\documentclass{article}
\newtheorem{theorem}{Theorem}

\begin{document}

\title{A NEW\ INTEGRABLE PROBLEM WITH A\ QUARTIC\ INTEGRAL IN THE DYNAMICS
OF A RIGID BODY}
\author{\textit{H. M. Yehia\thanks{%
E-mail:hyehia@mans.edu.eg} and A.A. Elmandouh\thanks{%
E-mail:adel78@mans.edu.eg}} \\
\textit{Department of Mathematics, Faculty of Science,}\\
\textit{\ Mansoura University, Mansoura 35516, Egypt}}
\maketitle

\begin{abstract}
We consider the problem of motion of a rigid body about a fixed point under
the action of an axisymmetric combination of potential and gyroscopic
forces. We introduce a new integrable case, valid on zero level of the
cyclic integal, that generalizes the known case of motion of a body in
liquid due to Chaplygin and its subsequent generalization by Yehia. Apart
from certain singular potential terms, the new case involves finite
potential and gyroscopic forces, which admit physical interpretation as
resulting from interaction of mass, magnetized parts and electric charges on
the body with gravitational, electric and magnetic fields.
\end{abstract}

\section{\label{intr}Introduction}

Consider a general problem of motion of a rigid body about fixed point $O$
under the action of a combination of conservative potential and gyroscopic
forces with a common axis of symmetry, the $OZ$-axis fixed in space. This
problem is described by the Lagrangian 
\begin{equation}
L=\frac{1}{2}\mathbf{\omega I\cdot \omega +l\cdot \omega -}V  \label{L}
\end{equation}%
in which $\mathbf{I=}diag(A,B,C)$ is the inertia matrix at $O,$ \bigskip $%
\mathbf{\omega }=(p,q,r)$ is the angular velocity of the body and $\mathbf{%
\gamma }=(\gamma _{1},\gamma _{2},\gamma _{3})$ is the unit vector in the
direction of the $Z$-axis, the scalar and vector potentials $V$ and $\mathbf{%
l}$, depend only on $\gamma _{1},\gamma _{2},\gamma _{3}$ and all vectors
are referred to the body system which we take as the system of principal
axes of inertia. Equations of motion\ in terms of the Euler-Poisson
variables $\mathbf{\omega }$ and $\mathbf{\gamma }$ are 
\begin{equation}
\mathbf{\dot{\omega}I+\omega \times (\omega I+\mu )=\gamma \times }\frac{%
\partial V}{\partial \mathbf{\gamma }},\qquad \mathbf{\dot{\gamma}}+\mathbf{%
\omega }\times \mathbf{\gamma }=\mathbf{0}  \label{ep}
\end{equation}%
where 
\begin{equation}
\mathbf{\mu }=(\mu _{1},\mu _{2},\mu _{3})=\frac{\partial }{\partial \mathbf{%
\gamma }}(\mathbf{l}\cdot \mathbf{\gamma )-(}\frac{\partial }{\partial 
\mathbf{\gamma }}\cdot \mathbf{l)\gamma }  \label{mu}
\end{equation}

Equations (\ref{ep}) admit three general first integrals: Jacobi's integral $%
I_{1}=\frac{1}{2}\mathbf{\omega I\cdot \omega +}V$, the geometric integral $%
I_{2}=\gamma _{1}^{2}+\gamma _{2}^{2}+\gamma _{3}^{2}=1$ and the cyclic
integral corresponding to cyclic angle of precession around the axis of
symmetry of the fields:%
\begin{equation}
I_{3}=(\mathbf{\omega I+l})\cdot \mathbf{\gamma }  \label{i3}
\end{equation}%
For those equations to be integrable, there should exist a fourth integral $%
I_{4}$, functionally independent of the above three. A case is named
"general" or "conditional" integrable according to whether this
complementary integral is valid for arbitrary initial conditions of the
motion or only on a single level of the integral $I_{3}$.

The problem (\ref{ep}) was formulated in \cite{yjmta86-1} and was addressed
in several subsequent works, mainly from the point of view of integrability.
This problem (\ref{ep}) represents a wide generalization of many classical
problems in rigid body dynamics that used to be considered separately.

\begin{enumerate}
\item \textbf{The classical problem of motion of a heavy rigid body about a
fixed} point in a uniform gravity field\ corresponds to the choice $\mathbf{%
l=\mu =0,}V=\mathbf{a\cdot \gamma }$ ($\mathbf{a}$ is a constant vector).
This is historically the most intensively studied version. For it we have
the famous three (and no more) general integrable cases known after Euler,
Lagrange and Kowalevski and one conditional integrable case on the zero
level of the integral $I_{3}$, bearing the names of Goriachev and Chaplygin
(see e.g. \cite{lei}).

\item \textbf{The problem of motion of a gyrostat}, which is a heavy rigid
body with a rotor spinning with a constant angular speed about its axis of
symmetry, fixed in the body. The gyrostatic moment can also be due to
internal cyclic degrees of freedom such as circulation of fluid in holes
inside the body or to forced stationary motions as motors and flow of fluids
in circuits in the body (see e.g. \cite{bormam}). In an interesting
alternative, due to Levi-Civita \cite{lc}, the rotor is left to move freely
around its axis of symmetry fixed in the body. In that case the matrix $%
\mathbf{I}$ is not simply the matrix of inertia of the system, but depends
on the direction of the rotor in the body and on the cyclic constant of its
motion.

This version corresponds to the choice $\mathbf{l=}$ $\mathbf{\mu =k\ }$($%
\mathbf{k}$ is a constant vector)$V=\mathbf{a\cdot \gamma .}$ For it we have
three general integrable cases named after Lagrange, Joukovsky (see e.g.\cite%
{lei}) and Yehia \cite{ymrc86}, which add the gyrostatic moment to the
integrable cases of problem 1. Gavrilov \cite{gav} has shown that no more
general integrable cases exist. The condtional case obtained by Sretensky 
\cite{sr}\ generalizes the above case of problem 1 due to Goriachev and
Chaplygin.

\item \textbf{The problem of motion of a rigid body by inertia in an ideal
incompressible fluid}, infinitely extending and at rest at infinity is
traditionally described for a simply connected body by Kirchhoff's equations 
\cite{kir70} (see also \cite{kir74}) or by their Hamiltonian form, mostly
used by mathematicians, which are due in their final form to Clebsch \cite%
{cleb}. For a perforated body (a body bounded by a multi-connected surface)
the equations of motion are usually taken in the form due to Lamb \cite{lam}%
, or in the equivalent Hamiltonian form (see e.g. \cite{bormam}). If one
writes those equations in the frame of reference attached to the principal
axes of a matrix $\mathbf{I}$ (the inertia matrix of the body modified with
an attached distribution of mass that compensates the presence of the
fluid), they involve 21 parameters characterizing the shape of the body and,
for a perforated body, circulations of the fluid along irreducible contours
on its surface. As discussed in \cite{yjmta86-2}, the traditional equations
of Kirchhoff and Lamb suffer some disadvantages that in most cases lead to a
their treatment in isolation from other problems of rigid body dynamics.

The problem under consideration has six degrees of freedom: three for the
rotational motion and three for the translation of a point of the body. A
new form of the equations of motion of a general body in a liquid was
derived by Yehia in \cite{yjmta86-2}. By eliminating the translational
motion of the body in terms of the angular velocity and a vector (constant
in space), the problem is reduced to one of three degrees of freedom. This
form completely fits in the system (\ref{ep}) of a body with one fixed point
and corresponds to the choice%
\begin{eqnarray}
\mathbf{l} &\mathbf{=}&\mathbf{k+\gamma K}  \nonumber \\
\mathbf{\mu } &\mathbf{=}&\mathbf{k+\gamma \bar{K},(\bar{K}=}\frac{1}{2}tr%
\mathbf{(K)\delta -K)}  \nonumber \\
V &=&\mathbf{a\cdot \gamma +}\frac{1}{2}\mathbf{\gamma J\cdot \gamma }
\label{bl}
\end{eqnarray}%
where $\mathbf{K,J}$ are constant symmetric $3\times 3$ matrices, $\mathbf{%
\delta }$ is $3\times 3$ unit matrix and $\mathbf{k,a}$ are constant
vectors. In this formulation the vectors $\mathbf{k,a,}$ which result from
the circulation of the fluid in the body perforations may be interpreted as
a gyrostatic moment and the centre of mass, respectively, of the equivalent
body moving about a fixed point.

The equations of motion (\ref{ep}) with $\mathbf{\mu }$ and $V$ in (\ref{bl}%
) involve only $18$ parameters. The three arbitrary parameters can always be
added to the expression for the translational motion of the central point of
the body.

For the last problem we know six general integrable cases, a list of which,
up to 1986, was provided in \cite{yjmta86-2}. The present number of general
integrable cases is still the same. The only change is in the last case
(case 6 of \cite{yjmta86-2}), which is now replaced by a recent case due to
Yehia \cite{yrcd03}. The latter is a one parameter generalization of an
earlier result of Sokolov \cite{sok}, which, in turn, was a one parameter
generalization of Yehia's gyrostat \cite{ymrc86}.

The list of conditional integrable cases of the present problem is composed
of two cases:

a) A case found by Yehia in \cite{ymrc87-III}, which generalizes a classical
result of Chaplygin \cite{ch903} by introducing five parameters and a later
result of Goriachev \cite{gor16} by three parameters.

b) A case found (with an additional singular potential term) in \cite%
{sok-tsig} generalizing the previous Sretensky and Goriachev-Chaplygin cases.

\item \textbf{The problem of motion of electrically charged rigid body }

\bigskip The potential $V$ can be understood in many cases as due to the
scalar interactions of a gravitational field with the mass distribution in
the body, an electric field with a permanent distribution of electric
charges and a magnetic field with some magnetized parts or steady currents
in electric circuits on the body. A constant term $\mathbf{k}$ of the
vectors $\mathbf{\mu }$ and $\mathbf{l}$ is a gyrostatic moment while the
variable terms may appear as a result of the Lorentz effect of the magnetic
field on the electric charges. For such a model to be realistic one must
also assume that the velocity of all mass elements and accelerations of
electric charges are sufficiently small to neglect both relativistic effects
and classical radiation damping. \smallskip Let \textbf{$\mathcal{B}$} and $%
\mathcal{A}$ be the intensity of the magnetic field and the vector potential
of this field at the point $\mathbf{r}$ of the body where the current charge
element $de$ is placed. In that case one can write the vector $\mathbf{l}$
as (for details see \cite{yjmta86-1})\footnote{%
Here MKS units are used. In Gaussian units $de$ should be devided by the
velocity of light $c$ (e.g. \cite{brad}).}:%
\[
\mathbf{l}=\mathbf{k}+\int \mathbf{r}\times \mathcal{A}de 
\]%
while $\mathbf{\mu }$ can be derived from $\mathbf{l}$ according to (\ref{mu}%
) or constructed directly in the form \cite{yjmta86-1}: 
\begin{equation}
\mathbf{\mu }=\mathbf{k}-\int (\mathbf{r}\cdot \mathbf{\mathcal{B}})\mathbf{r%
}de  \label{mud}
\end{equation}%
It was noted in \cite{yjmta86-1}, that the variable parts of the vector $%
\mathbf{\mu }$ appear also in the case of a moving dielectric body in a
combination of electric and magnetic fields.

As all the forces acting on the body are supposed to be symmetric about the $%
Z$-axis, the combination of static external fields: gravitational field with
potential $\Phi _{g}$, electric field of potential $\Phi _{e}$ and magnetic
field $\mathcal{B}$ whose scalar potential is $\Phi $ can depend only on $Z$
and $X^{2}+Y^{2}$. The potential of the body can be written as%
\begin{equation}
V=\int (\Phi _{g}dm+\Phi _{e}de+\mathcal{B}\cdot d\mathbf{\sigma ),}
\label{V}
\end{equation}%
where $dm,de$ and $d\mathbf{\sigma }$ are the mass, electric charge and the
magnetic moment contained in the element of the body which at the current
moment occupies the point $\mathbf{r}(X,Y,Z)$ of the inertial frame. In most
cases we find it is suitable to choose the three potentials $\Phi _{g},\Phi
_{e}$ and $\Phi $ to be polynomials in $X,Y,Z$, subject to Laplace's
equation and to the axial symmetry condition. Due to the abundance of
physical parameters representing the three distributions and the
coefficients of the three potentials, it should be easy to adjust those
parameters to match the potential in each case and, moreover, in a variety
of choices.

Examples of integrable cases of the present problem are the first four
general integrable cases introduced in \cite{yjpa-mu}. In those cases the
potential and the components of $\mathbf{\mu }$ are all polynomials in the
Poisson variables $\gamma _{1},\gamma _{2},\gamma _{3}$. Two of those cases
were given interpretations in the frame of the present problem in \cite%
{yjpaall}.
\end{enumerate}

From the mathematical point of view, each of the above problems is a special
case of the one coming after it and integrable cases of one problem are
usually generalizable to the later ones by means of adding extra-parameters
describing additional physical effects. However, it turned out that some
integrable potential terms are not likely to be counted for in any of the
above four frames. Examples are the potential term $\frac{1}{\gamma _{3}^{2}}
$ introduced by Goriachev \cite{gor16} and the term $\frac{1}{\gamma _{3}^{4}%
}-\frac{1}{\gamma _{3}^{6}}$ found by Yehia in \cite{yrcd03}. Singular terms
occur also in several other cases integrable on the zero level of the linear
integral $I_{3}$ (e.g. \cite{yrcd06}, \cite{ymJPA11}). Terms, singular on
the plane through the fixed point perpendicular to the axis of symmetry of
the fields, cannot be explained as due to electric or magnetic forces, since
the latter can have only singular points of the Coulomb type. Thus, certain
generalizations of the classical integrable cases are bound to the general
problem \ref{ep} and not to any of the four physical problems.

In the present note we announce a new conditional case, valid on the level $%
I_{3}=0.$The new case adds one parameter to the structure of a previous
result of Yehia \cite{yrcd03}. It is also a 4-parameter generalization of
the well-known case of Chaplygin. in the dynamics of a rigid body moving by
inertia in an infinitely extended ideal incompressible fluid \cite{ch903}.
Unlike most previous generalizations of integrable cases in rigid body
dynamics, the new parameter introduced here evokes finite potential and
gyroscopic terms in the equations of motion. This situation makes it
possible to construct a physical interpretation of the finite terms as
resulting from gravitational, electric and Lorentz interactions.

\section{A new integrable problem}

The method devised in 1986 by Yehia \cite{yint-86} has proved highly
effective for constructing 2D conservative mechanical systems whose
configuration spaces are, in general, Riemannian manifolds, which admit a
complementary first integral polynomial in the velocity variables. It is
suitable for time-reversible systems involving only forces of potential
character and also for generalized-natural systems, involving gyroscopic
forces as well. For the last systems, the Lagrangian functions include terms
linear in velocities.

The culmination of this method for time-reversible systems was the
construction of the so-called "master" system \cite{yjpa-master}, which
admits a quartic integral and involves the unprecedented number of $21$
arbitrary parameters in its structure. This system includes as special cases
of it, corresponding to certain values of the parameters, two new integrable
cases of rigid body dynamics. Two systems with a cubic integral were found
in \cite{yjpa-cu}. One of them is a many-parameter generalization of the two
well-known cases of integrable rigid body dynamics, known after Goriachev
and Chaplygin \cite{lei} and Goriachev \cite{gor16}.

The situation as to time-irreversible systems is much harder. To date, the
above mentioned method has produced 41 irreversible cases with a quadratic
integral \cite{yjpa-92}, \cite{yjmp-07}. All relevant known integrable cases
of motion of rigid bodies are restored as special cases of some of the new
systems and new ones were constructed \cite{yjmp-07}. The general system
with a cubic integral of \cite{yjpa-cu} is generalized into an irreversible
system by adding two parameters. This also induced a generalization of the
cases of Goriachev and Chaplygin and Goriachev by adding potential and
gyroscopic forces, which preserves integrability.

\bigskip The PDEs that determine gyroscopic generalizations of of systems
with a quartic integral follow from \cite{yint-86} and are written in detail
in \cite{ym-rcd08}. Unlike the case of cubic integrals, they have not been
solved for general arbitrary values of parameters in the reversible system.
A solution of those equations is found in \cite{ym-rcd08}, corresponding to
the parameter values that lead only to rigid body dynamics. A new similar
version of the same problem is obtained using the same method with a
different ansatz of the solution. The resulting Lagrangian system can be
written in terms of Euler's angles as generalized coordinates, but the form
of the integral is not much tractable. Instead, we present here the final
result as an explicit case of integrable Euler-Poisson equations. In this
traditional formalism of rigid body dynamics it is easy to check the
correctness of the integral and also to compare the present result with the
previously known integrable cases of the problem. Thus we formulate the
following

\begin{theorem}
Let the moments of inertia satisfy the Kowalevski condition $A=B=2C$\ and
let the scalar and vector functions $V$ and $\mu $\ be given by%
\begin{eqnarray}
V &=&C\{\kappa \left[ 2d\gamma _{1}\gamma _{2}+c(\gamma _{1}^{2}-\gamma
_{2}^{2})\right] +\frac{1}{2}n^{2}\gamma _{3}^{2}-nK\gamma _{3}[d(\gamma
_{2}^{2}-\gamma _{1}^{2})+2c\gamma _{1}\gamma _{2}]  \nonumber \\
&&+K^{2}\left[ 2cd\gamma _{1}\gamma _{2}(\gamma _{1}^{2}-\gamma _{2}^{2})+%
\frac{d^{2}}{2}(\gamma _{3}^{4}+4\gamma _{1}^{2}\gamma
_{2}^{2})-c^{2}(\gamma _{3}^{2}(\gamma _{1}^{2}+\gamma _{2}^{2})+2\gamma
_{1}^{2}\gamma _{2}^{2})\right]  \nonumber \\
&&+\frac{\lambda }{\gamma _{3}^{2}}+\rho (\frac{1}{\gamma _{3}^{4}}-\frac{1}{%
\gamma _{3}^{6}})\}.  \label{pot}
\end{eqnarray}%
and%
\begin{equation}
\mathbf{\mu }=C\left( 2K\gamma _{3}(c\gamma _{2}-d\gamma _{1})-n\gamma
_{1},~2K\gamma _{3}(d\gamma _{2}+c\gamma _{1})-n\gamma _{2},~K[d(\gamma
_{2}^{2}-\gamma _{1}^{2})+2c\gamma _{1}\gamma _{2}]-3n\gamma _{3}\right)
\label{muu}
\end{equation}%
or, equivalently, 
\[
\mathbf{l}=C(2n\gamma _{1},2n\gamma _{2},n\gamma _{3}+K[d(\gamma
_{2}^{2}-\gamma _{1}^{2})+2c\gamma _{1}\gamma _{2}]) 
\]%
where $c,d,n,\lambda ,\rho ,\kappa $~and $K$ are free parameters, then
Euler-Poisson equations of motion (\ref{ep}), which in the present case take
the form%
\begin{eqnarray*}
\dot{p} &=&\frac{1}{2}qr-\frac{q}{2}\left[ K(d(\gamma _{2}^{2}-\gamma
_{1}^{2})+2c\gamma _{1}\gamma _{2})-3n\gamma _{3}\right] +\frac{r}{2}\left[
2K(c\gamma _{1}+d\gamma _{2})-n\gamma _{2}\right] \\
&&+K^{2}\gamma _{3}\left[ cd\gamma _{1}(3\gamma _{2}^{2}-\gamma
_{1}^{2})-d^{2}\gamma _{2}(2\gamma _{1}^{2}-\gamma _{3}^{2})+c^{2}\gamma
_{2}(\gamma _{1}^{2}+\gamma _{3}^{2}-\gamma _{2}^{2})\right] \\
&&+\kappa \gamma _{3}(c\gamma _{2}-d\gamma _{1})-\frac{nK}{2}[2c\gamma
_{1}(\gamma _{2}^{2}-\gamma _{3}^{2})-\gamma _{2}d(2\gamma _{3}^{2}-\gamma
_{2}^{2}+\gamma _{1}^{2})] \\
&&+\gamma _{2}\left( \frac{n^{2}\gamma _{3}}{2}-\frac{\lambda }{\gamma
_{3}^{3}}-\frac{2\gamma _{3}^{2}-3}{\gamma _{3}^{7}}\rho \right) , \\
\dot{q} &=&-\frac{pr}{2}+\frac{p}{2}\left[ K(d(\gamma _{2}^{2}-\gamma
_{1}^{2})+2c\gamma _{1}\gamma _{2}\right] -\frac{r}{2}\left[ 2K(c\gamma
_{2}-d\gamma _{1})-n\gamma _{1}\right] \\
&&-K^{2}\gamma _{3}\left[ cd\gamma _{2}(3\gamma _{1}^{2}-\gamma
_{2}^{2})-d^{2}\gamma _{1}(2\gamma _{2}^{2}-\gamma _{3}^{2})+c^{2}\gamma
_{1}(\gamma _{2}^{2}+\gamma _{3}^{2}-\gamma _{1}^{2})\right] \\
&&+\kappa \gamma _{3}(c\gamma _{1}+d\gamma _{2})+\frac{nK}{2}\left[ 2c\gamma
_{2}(\gamma _{1}^{2}-\gamma _{3}^{2})-d\gamma _{1}(\gamma _{1}^{2}-\gamma
_{2}^{2}-2\gamma _{3}^{2})\right] \\
&&+\gamma _{1}\left( \frac{2\gamma _{3}^{2}-3}{\gamma _{3}^{7}}\rho +\frac{%
\lambda }{\gamma _{3}^{3}}-\frac{n^{2}}{2}\gamma _{3}\right) , \\
\dot{r} &=&\left[ 2K(c\gamma _{2}-d\gamma _{1})\gamma _{3}-n\gamma _{1}%
\right] q-\left[ 2K(c\gamma _{1}+d\gamma _{2})\gamma _{3}-n\gamma _{2}\right]
p \\
&&+2K^{2}\left[ cd(\gamma _{1}^{4}+\gamma _{2}^{4}-6\gamma _{1}^{2}\gamma
_{2}^{2})-2(c^{2}-d^{2})\gamma _{1}\gamma _{2}(\gamma _{1}^{2}-\gamma
_{2}^{2})\right] \\
&&-2\kappa \lbrack 2c\gamma _{1}\gamma _{2}+d(\gamma _{2}^{2}-\gamma
_{1}^{2})]-2nK\gamma _{3}\left[ c(\gamma _{1}^{2}-\gamma _{2}^{2})+2d\gamma
_{1}\gamma _{2}\right] , \\
\dot{\gamma}_{1} &=&r\gamma _{2}-q\gamma _{3} \\
\dot{\gamma}_{2} &=&p\gamma _{3}-r\gamma _{1} \\
\dot{\gamma}_{3} &=&q\gamma _{1}-p\gamma _{2}
\end{eqnarray*}%
are integrable on the zero level of the integral%
\[
I_{3}=2p\gamma _{1}+2q\gamma _{2}+{\Large \{}r+K\left[ d(\gamma
_{2}^{2}-\gamma _{1}^{2})+2c\gamma _{1}\gamma _{2}\right] {\Large \}}\gamma
_{3}-n\gamma _{3}^{2} 
\]%
The complementary integral of the motion is%
\begin{eqnarray}
I_{4} &=&\left[ (p+n\gamma _{1})^{2}-(q+n\gamma _{2})^{2}+c\kappa \gamma
_{3}^{2}+\gamma _{3}^{2}\left( Kd(r+n\gamma _{3})+cK^{2}(c(\gamma
_{1}^{2}-\gamma _{2}^{2})+2d\gamma _{1}\gamma _{2}\right) -\frac{\lambda
(\gamma _{1}^{2}-\gamma _{2}^{2})}{\gamma _{3}^{2}}\right] ^{2}  \nonumber \\
&&+\left[ 2(p+n\gamma _{1})(q+n\gamma _{2})+d\kappa \gamma
_{3}^{2}+(dK^{2}(c(\gamma _{1}^{2}-\gamma _{2}^{2})+2d\gamma _{1}\gamma
_{2})-Kc(r+n\gamma _{3}))\gamma _{3}^{2}-\frac{2\lambda \gamma _{1}\gamma
_{2}}{\gamma _{3}^{2}}\right] ^{2}  \nonumber \\
&&+\rho \{\frac{2(\gamma _{3}^{2}-1)}{\gamma _{3}^{6}}[(p+n\gamma
_{1})^{2}+(q+n\gamma _{2})^{2}]-\frac{2K(r+n\gamma _{3})[2c\gamma _{1}\gamma
_{2}+d(\gamma _{2}^{2}-\gamma _{1}^{2})]}{\gamma _{3}^{4}}  \nonumber \\
&&+\frac{(1-\gamma _{3}^{2})^{2}}{\gamma _{3}^{12}}(\rho -2\lambda \gamma
_{3}^{4})+K^{2}[2c^{2}(\frac{1}{\gamma _{3}^{4}}-\frac{2}{\gamma _{3}^{2}})+8%
\frac{(d^{2}-c^{2})\gamma _{1}^{2}\gamma _{2}^{2}+cd\gamma _{1}\gamma
_{2}(\gamma _{1}^{2}-\gamma _{2}^{2})}{\gamma _{3}^{4}}]  \nonumber \\
&&+\frac{2\kappa }{\gamma _{3}^{4}}[c(\gamma _{1}^{2}-\gamma
_{2}^{2})+2d\gamma _{1}\gamma _{2}]\}.
\end{eqnarray}
\end{theorem}

This case, involving 7 significant parameters, is a new generalization of
the previously known integrable problems in rigid body dynamics:

\bigskip 
\begin{tabular}{|l|l|}
\hline
Author- year & Conditions on parameters \\ \hline
Yehia \cite{yrcd03} (\S 4.2.3) 2003 & $K=0$ \\ \hline
Goriachev\cite{gor16} 1916 & $K=n=\rho =0$ \\ \hline
Chaplygin \cite{ch903} 1903 & $K=n=\rho =\lambda =0$ \\ \hline
\end{tabular}

\section{Physical interpretation:}

\bigskip The Goriachev parameter $\lambda $ and Yehia's parameter $\rho $
give rise to a singular plane of the potential. That is the plane $\gamma
_{3}=0,$ orthogonal to the space axis of symmetry of the fields applied to
the body. For a physical interpretation we set 
\begin{equation}
\lambda =\rho =0  \label{con}
\end{equation}%
The potential becomes a quartic polynomial in the Poisson variables
containing cubic and quadratic terms. As all the forces acting on the body
are symmetric about $Z-$ axis. Let there be the following combination of
static external fields: a gravitational field with potential $\Phi _{g}$ ,
an electric field of potential $\Phi _{e}$ and magnetic field \textbf{$%
\mathcal{B}$} whose scalar potential is $\Phi $. Note that the three
potentials can depend only on $Z$ and $X^{2}+Y^{2}$. The potential of the
body can be written in the form (\ref{V}), which in the present case is a
polynomial expression of degree 4 in the components of $\mathbf{\gamma }$.
Thus, in this case we find that it suffices to choose the three potentials $%
\Phi _{g}$, $\Phi _{e}$ and $\Phi $ to be polynomial solutions of Laplace's
equation subject only to the axial symmetry condition. The potential $V$ can
be realized in a variety of ways.

The vector $\mathbf{\mu }$ can be expressed directly in terms of the
magnetic field according to the formula (\ref{mud}).

In the case when the scaler potential of the external magnetic field can be
expressed as a second-degree harmonic polynomial%
\begin{equation}
\Phi =a_{1}Z+a_{2}(3Z^{2}-r^{2})  \label{p3}
\end{equation}%
The vector $\mathbf{\mu }$ can be expressed in terms of the body system of
coordinates,%
\begin{equation}
\mu =\int [a_{1}\mathbf{r\cdot \gamma }+2a_{2}(3(\mathbf{r\cdot \gamma }%
)^{2}-r^{2})]\mathbf{r}de  \label{p4}
\end{equation}%
Finally, we can write%
\begin{eqnarray*}
\mu _{1} &=&-2a_{2}(I_{xxx}+I_{xyy}+I_{xzz})+a_{1}(I_{xx}\gamma
_{1}+I_{xy}\gamma _{2}+I_{xz}\gamma _{3}) \\
&&+6a_{2}(I_{xxx}\gamma _{1}^{2}+I_{xyy}\gamma _{2}^{2}+I_{xzz}\gamma
_{3}^{2}+2I_{xxy}\gamma _{1}\gamma _{2}+2I_{xxz}\gamma _{1}\gamma
_{3}+2I_{xyz}\gamma _{2}\gamma _{3})
\end{eqnarray*}%
\begin{eqnarray}
\mu _{2} &=&-2a_{2}(I_{xxy}+I_{yyy}+I_{yzz})+a_{1}(I_{xy}\gamma
_{1}+I_{yy}\gamma _{2}+I_{yz}\gamma _{3})  \label{p5} \\
&&+6a_{2}(I_{xxy}\gamma _{1}^{2}+I_{yyy}\gamma _{2}^{2}+I_{yzz}\gamma
_{3}^{2}+2I_{xyy}\gamma _{1}\gamma _{2}+2I_{xyz}\gamma _{1}\gamma
_{3}+2I_{yyz}\gamma _{2}\gamma _{3})  \nonumber
\end{eqnarray}%
\begin{eqnarray*}
\mu _{3} &=&-2a_{2}(I_{xxz}+I_{yyz}+I_{zzz})+a_{1}(I_{xz}\gamma
_{1}+I_{yz}\gamma _{2}+I_{zz}\gamma _{3}) \\
&&+6a_{2}(I_{xxz}\gamma _{1}^{2}+I_{yyz}\gamma _{2}^{2}+I_{zzz}\gamma
_{3}^{2}+2I_{xyz}\gamma _{1}\gamma _{2}+2I_{xzz}\gamma _{1}\gamma
_{3}+2I_{yzz}\gamma _{2}\gamma _{3})
\end{eqnarray*}%
where, for example, $I_{xx}=\int x^{2}de$, $I_{xyz}=\int xyzde$ and so forth
are the second- and third-degrees moments of the charge distribution.

It is not hard now to verify that the finite-potential version the new
integrable case, subject to the conditions (\ref{con}), corresponds to the
choice%
\begin{eqnarray}
I_{xy}
&=&I_{xz}=I_{yz}=I_{xxx}=I_{xxy}=I_{xyy}=I_{xzz}=I_{yyy}=I_{yzz}=I_{zzz}=I_{xxz}+I_{yyz}=0
\nonumber \\
I_{yy} &=&I_{xx}=\frac{I_{zz}}{3}=-\frac{Cn}{a_{1}},%
6a_{2}I_{xxz}=-CKd,6a_{2}I_{xyz}=cCK  \label{p6}
\end{eqnarray}%
This is a set of conditions on the second and third moments of the charge
distribution. Regarding the fact that \bigskip the charge distribution is
not necessarily positive, no restrictions like the triangle inequalities on
inertia moments are necessary here.\ It turns out that the charge
distribution can be determined with great arbitrariness.

\end{document}